\documentclass[pre,manuscript,showpacs]{revtex4}

\usepackage{graphicx}
\usepackage{times}
\usepackage{color}

\usepackage{amssymb,amsfonts,amsmath}

\begin{document}

\title{\textbf{\huge Evolutionary games on minimally
structured populations}}
\author{Gergely J. Sz\"oll\H{o}si}
\email[]{ssolo@angel.elte.hu}
\homepage[]{angel.elte.hu/~ssolo}
\author{Imre Der\'enyi}
\email[]{derenyi@angel.elte.hu}
\homepage[]{angel.elte.hu/~derenyi}
\affiliation{Biological Physics Department E\"otv\"os University, Budapest}

\begin{abstract} 

Population structure induced by both spatial embedding and more general networks of interaction, 
such as model social networks, have been shown to have a fundamental effect on the dynamics and outcome of 
evolutionary games. These effects have, however, proved to be sensitive to the details of the underlying topology 
and dynamics. Here we introduce a minimal  
population structure that is described by two distinct hierarchical levels of interaction, 
similar to the structured metapopulation concept of ecology and island models in population genetics. 
We believe this model is able to identify    
effects of spatial structure that do not depend on the details of the topology. 
While effects depending on such details clearly lie outside the scope of our approach, we expect that 
those we are able to reproduce should be generally applicable to a wide range of models.
We derive the dynamics governing the evolution of a 
system starting from fundamental individual level 
stochastic processes through two successive meanfield approximations. 
In our model of population structure the topology of interactions 
is described by only two parameters: the effective population 
size at the local scale and the relative strength of 
local dynamics to global mixing.  We demonstrate, for example, the existence of a continuous transition leading to the dominance of cooperation in populations with  hierarchical levels of unstructured mixing as the benefit to cost ratio becomes smaller then the local population size. Applying our model of spatial structure to 
the repeated prisoner's dilemma we uncover a novel and counterintuitive mechanism by which the constant influx of defectors sustains cooperation.    
Further exploring the phase space of the repeated prisoner's dilemma 
and  also of 
the  ``rock-paper-scissor'' game we find indications of rich structure and are able to reproduce 
several effects observed in other models with explicit spatial embedding, such as the maintenance of biodiversity and
the emergence of global oscillations.
\end{abstract}
\maketitle

\section{Introduction}
The dynamics of Darwinian evolution is intrinsically
frequency dependent, the fitness of individuals is tightly coupled to
the type and number of competitors. Evolutionary dynamics acts, however,
on populations, not individuals and as a consequence depends on not only
population composition, but also population size and structure. 
Evolutionary game theory came about as the result of the realization that frequency dependent fitness
introduces strategic aspects to
evolution \cite{Fisher,JMS,HofbauerBOOK}. More recently the
investigation of the evolutionary dynamics of structured populations, where
individuals only compete with some subset of the population, e.g.\ 
their neighbors in space or more generally in some
graph \cite{NowakNAT92,LiebermanNAT}, 
has lead to the recognition that the success of different strategies
can be greatly influenced by the topology of interactions within the
population.
Fundamental differences were found -- compared to well-mixed populations, where individuals interact with
randomly chosen partners --  in models that describe the evolution of cooperation
(variants of the prisoner's dilemma
game  \cite{NowakNAT92,Axelrod,AxelrodBOOK,NowakPNAS94,HauertPRL02}) or
deal with the maintenance of biodiversity in the context of
competitive cycles (variants of the rock-paper-scissors
game  \cite{HofbauerBOOK,KerrNAT,NowakNATnw,KirkupNAT,CzaranPNAS,LenskiPNASComm}). 

In order to investigate the coevolutionary dynamics of games on structured populations
the full set of connections between a
potentially very large number of individuals must be specified. This is only possible by
reducing the number of degrees of freedom considered, either through
postulating a highly symmetrical (such as lattices  \cite{NowakNAT92,NowakPNAS94,SzaboPRE98,SzaboClique,GTphy,Hui07,Hui05}) or fundamentally
random connection structure  (such as some random graph ensemble \cite{SantosPRL,OhtsukiNAT06}). The
question of how one goes about the task of reducing the number of
degrees of freedom -- of choosing the relevant  parameters to describe
the population structure constrained to which individuals undergo
evolution --  is not trivial. Both the explicit spatial as well as the
random graph ensemble approach have clear precedents  in condensed
matter physics and network theory, respectively. It is not, however,
clear which -- if either -- approach best describes natural
populations of cyclically competing species or societies composed of
individuals playing the prisoner's dilemma game. 

As an example let us consider colicin producing bacteria, that play the so called
"rock-paper-scissors" (RPS) game (for details see below). This system has
recently been the subject of two experimental studies aimed at
demonstrating the role of structured populations in the maintenance of
diversity. In the first study  \cite{KerrNAT,NowakNATnw} bacteria were cultured \emph{in
vitro} in Petri dishes, effectively restricting competition between
bacteria to neighbors on the (2D) Petri dish surface (Fig.1 top left),
while in
the second experiment  \cite{KirkupNAT} \emph{in vivo} bacterial
colonies were established in co-caged mice and their development was
subsequently followed. In the case of the first experiment the analogy
with explicit 2D spatial embedding (present by construction) is clear
(Fig.1 bottom left). The population structure of the second experiment
is, however, clearly different. The bacteria in individual mice can be
readily considered as locally well-mixed populations, the
coevolutionary dynamics of which reduces in the standard meanfield
limit to a system of non-linear differential equations (the
adjusted replicator equations  \cite{TraulsenPRL}). As the experiments
show, however, migration of bacteria between mice may also occur --
resulting in the observed cyclic presence of the three strains in
individuals. There are two distinct scales of mixing present in
the system. Bacteria within each mice compete with each other forming
local populations  -- an unstructured neighborhood composed of individual
bacteria, while also being exposed to migrants from mice with whom
they share the cage, together forming a global population -- an unstructured
neighborhood composed of individual local populations
(Fig.1 top and bottom right). This setup is referred to in the ecology literature
 -- albeit in significantly different contexts -- as a ''structured metapopulation''
  \cite{MetapopBOOK,Hanski} where structured here refers to the
detailed consideration of the population dynamics of the individual
populations (often called ''patches'') comprising the metapopulation and is also  related to the finite island models of population genetics \cite{Pannell}.

The above example of co-caged mice is not unique, we may readily think of other
ecological or sociological examples where an approximation with
hierarchical scales of mixing with no internal structure can be
relevant (such as human societies with two
distinct scales of mixing present, the first within individual nations
the between them at an international level). We have, also, 
recently used a similar approach to construct a model of genetic exchange among
bacteria of the same species (the bacterial equivalent of sex) with
which we were able to take into account the effects of spatial and
temporal fluctuations in a manner that can explain the benefit of such
genetic exchange at the level of the individual  \cite{SzollosiGEN}. 

In this paper we construct a hierarchical meanfield theory where the
two distinct (i.e.\ local and global) scales of mixing are each taken into account in terms of two
separate {\it meanfield} approximations and fluctuations 
resulting from finite population size on the local scale of mixing are
also considered.
We subsequently explore the similarities and differences between this 
and other models of structured populations in the case of the 
''rock-paper-scissors'' and prisoner's dilemma games. Through
these examples we
suggest that our
approach allows the separation of the effects of
structured populations on coevolutionary dynamics into effects which
are highly sensitive to and dependent on the details of the topology
and those which only require the minimal structure present in our
approximation and can consequently (in terms of sensitivity to the
details of the topology) be considered more robust.

\begin{figure}
\begin{center}
\centerline{\includegraphics[width=
.6\textwidth]{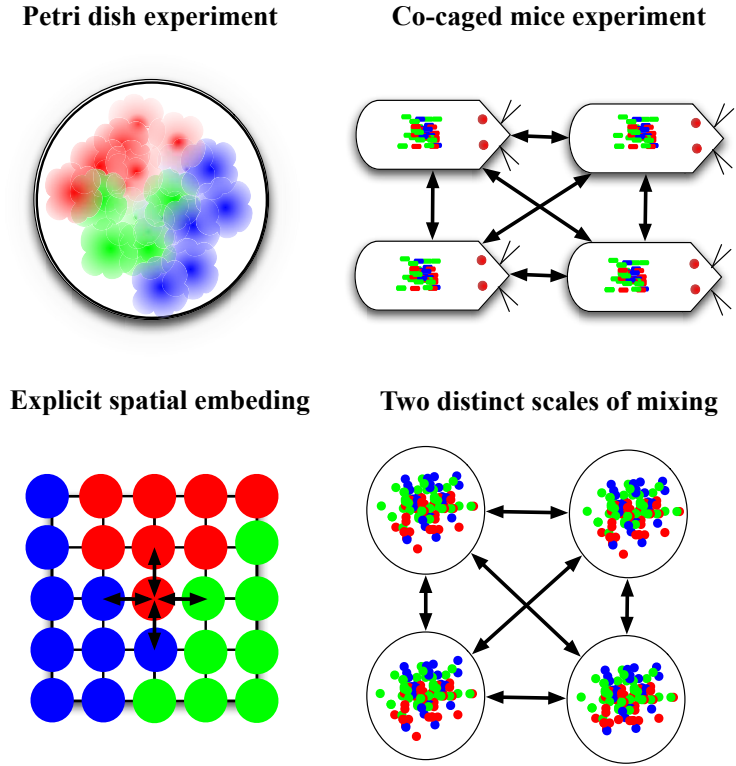}}
\caption{ (Color
 online) In the colicin version of the RPS game, strains that produce
colicins (red/dark grey) kill sensitive (green/light grey) strains, that outcompete
resistant (blue/black) strains, that outcompete colicin producing strains
(toxin production involves bacterial suicide).
Experiments  \cite{KerrNAT} show that colicin-producing strains cannot
coexist with sensitive or resistant strains in a well-mixed culture,
yet all three phenotypes are recovered in natural populations. Two
recent experiments have examined the role of population structure in
the maintenance of diversity among colicin-producing bacteria. 
In the first  \cite{KerrNAT} \emph{in vitro} colonies were established 
on an agar substrate in Petri dishes, a setup which effectively 
limits competition to neighbors on the petri dish in analogy with explicit spatial 
embedding in 2D. In the second  \cite{KirkupNAT} {\it in vivo} 
colonies were established in the intestines of co-caged mice, 
a setup which has two distinct scales of mixing, with no explicit 
structure on either scale.  
}\label{fig1}
\end{center}
\end{figure}

\section{ Hierarchical Meanfield Theory for Two Distinct Scales}    
Let us consider an evolutionary game between $d$ types (strategies) described by the 
$d\times d$ payoff matrix $A$ with elements $\alpha_{kj}$. 
Assuming finite and constant population size,  natural
selection can be described at the level of the individual by the so called the Moran
process  \cite{Moran},  during which at each time step an individual is selected
randomly from the population to be replaced (death) by the offspring of an individual  that is chosen proportional to its fitness to reproduce (birth). This models a population in equilibrium, where the time scale of the population dynamics is set by the rate at which ''vacancies'' become available in the population.  The fitness of each individual
 depends on the payoff received from playing the game described
by $A$ with competitors (an individual of type $k$ receiving a payoff $\alpha_{kj}$ when playing with an individual of type $j$). In well-mixed populations, individuals can be
considered to come into contact (compete) with equal probability with any member of the
population excluding themselves -- this allows one to calculate the
fitness of an individual of type $k$ in a meanfield manner, yielding 
\begin{equation}
\pi_k = \pi_{\rm base}+ \sum_{j=1}^{d} \frac{\alpha_{kj} (n_j-\delta_{kj})}{N-1},
\label{pis}
\end{equation}
where $n_k$ is the number of individuals of type $k$ in the
population, $\sum_{k=1}^{d} n_k = N$ is the size of the population, $\pi_{\rm base}$ is some baseline
fitness and the Kronecker delta symbol $\delta_{kj}$ is equal to unity if $k=j$ and is zero otherwise. From this we may calculate the transition probabilities of
our stochastic process, i.e., the probability of an individual of
type $i$ being replaced by an offspring of an individual of type $k$ is given by
\begin{equation}
T_{ik} =\frac{n_i}{N}\frac{\pi_k n_k }{\bar\pi N},\label{Ts}
\end{equation} 
where $\bar\pi=\sum_{k=1}^{d} \pi_k n_k/ N$.
The state of any population is completely described by the frequency of the different strategies $x_k=n_k/N$. Due to the normalization $\sum_{k=1}^N x_k =1$, the values of $x_k$ are restricted to the unit simplex $S_d$  \cite{HofbauerBOOK}. For $d=2$ this is the interval $[0,1]$, $S_3$ is the triangle with vertices $\{ (1,0,0),(0,1,0),(0,0,1) \}$ while $S_4$ is a tetrahedron etc.

As Traulsen {\it et al.\ }have recently shown  \cite{TraulsenPRL,TraulsenPRE} for sufficiently
 large, but finite populations the above stochastic process can be well approximated  by a set of stochastic differential equations combining deterministic dynamics and diffusion (population drift) referred to as Langevin dynamics:
\begin{equation}
\dot x_k = a_k({\bf x}) +  \sum_{j=1}^{d-1} c_{kj} ({\bf x}) \xi_j(t),
\label{single_langevin_eq}
\end{equation} 
where the effective deterministic terms $a_k({\bf x})$ are given by 
\begin{equation}
a_k({\bf x})=\sum_{j=1}^d(T_{jk}-T_{kj})=x_k\frac{\pi_k({\bf x})-\bar\pi({\bf x})}{\bar\pi({\bf x})} , 
\label{ak}
\end{equation} 
$c_{kj}({\bf x})$ are effective 
diffusion terms, that can also be expressed in terms of the transition
probabilities as described in  \cite{TraulsenPRE}, and $\xi_j$ are delta correlated
$\langle\xi_k(t)\xi_j(t')\rangle=\delta_{kj}\delta(t-t') $ Gaussian
white noise terms. As $N\to\infty$ the diffusion term
tends to zero as $1/ \sqrt{N}$ and we are left with the modified replicator equation.

In the context of our hierarchical mixing model the topology of
connections can be 
described by two parameters, the populations size at the local scale of mixing $N$, 
and a second parameter $\mu$, which tunes the strength of global mixing relative to
the local dynamics.  
We take into account the second (global) scale
of mixing -- mixing among local populations -- by introducing a modified
version of the 
Moran process.
In the modified process a random individual is replaced at each time
 step either with the offspring of 
an individual from the same population (local reproduction) or with an individual from the
global population (global mixing).
This is equivalent to considering the global
population to be well-mixed at the scale of local populations.

Let us consider a global population that is composed of $M$ local populations of size $N$.
In each local population vacancies become available that local reproduction and global mixing compete to fill.
In any local population $l$ the probability of an individual of some type $k$ filling a new vacancy due to local reproduction must be proportional to the number of individuals of type $k$ multiplied by their fitness i.e.\ $\pi_k^l n_k^l$, where we consider $\pi_k^l$ to be determined only by interactions with individuals in the same local population according to equation (\ref{pis}). To describe the tendency of individuals of some type $k$ in local population $l$ to contribute to global mixing we introduce the parameters $\sigma_k^l$. The choice of appropriate $\sigma_k^l$ depends on the details of the global mixing mechanism, for systems where only the offspring of individuals mix globally it is proportional to the fitness of a given type, while for mechanisms such as physical mixing, by e.g.\ wind or ocean currents, it may be identical for each type. Irrespective of the  details,   however, the probability of an individual of some type $k$ filling in a new vacancy due to global mixing should be proportional to the global average of the number of individuals of type $k$ multiplied by their mixing tendency, which we 
 denoted as  $\langle \sigma_k n_k \rangle =\sum_{l=1}^{M} \sigma_k^l n_k^l / M $, and the strength of global mixing $\mu$. These consideration lead to the new transition probabilities: 

\begin{equation}
\hat T^l_{ik} =
\frac{n^l_i}{N}  \left( \frac{ \pi_k^l n_k^l + \mu \langle \sigma_k n_k \rangle }{ \sum_{k=1}^d (\pi_k^l n_k^l + \mu \langle \sigma_k n_k \rangle)} \right )
 =
\frac{n^l_i}{N}  \left( \frac{ \pi_k^l n_k^l + \mu \langle \sigma_k n_k \rangle }{ N ( \bar\pi^l + \mu \bar{\langle \sigma \rangle})} \right )
,\label{muTs}
\end{equation}     
where $\bar\pi^l = \sum_{k=1}^{d} \pi^l_k n^l_k/N$ and  $\bar{\langle \sigma \rangle}= \sum_{k=1}^d   \langle\sigma_k n_k \rangle/N$.

 We have found that the results presented below are qualitatively the same for  both the \emph{fitness dependent} choice of $\sigma_k^l=\pi_k^l$ and the  \emph{fitness independent} choice of $\sigma_k^l=1$. Therefore, in the following we restrict ourselves to the somewhat simpler \emph{fitness independent} choice of $\sigma_k^l=1$, which can be considered to correspond to some form of physical mixing mechanism. The transition probabilities  (\ref{muTs}) then reduce to:
\begin{equation}
\bar T^l_{ik} = \frac{n_i^l}{N} \left( \frac{\bar \pi^l}{\bar\pi^l+\mu}\frac{\pi_k^l n^l_k}{\bar \pi^l N}  + \frac{\mu}{\bar\pi^l+\mu}\frac{\langle n_k \rangle}{N}  \right).
\label{muTsi}
\end{equation}
We can see that after a vacancy appears either local reproduction occurs, with probability $\bar \pi^l/ (\bar\pi^l +\mu)$, or global mixing, with probability  $\mu/ (\bar\pi^l +\mu)$.
From (\ref{muTsi}) we may derive the Langevin equation describing the 
coevolutionary dynamics of population $l$ from the 
\begin{equation} 
\dot x^l_k =  \hat a_k({\bf x}^l, \langle {\bf x} \rangle) +
\sum_{j=1}^{d-1} \hat c_{kj} ({\bf x}^l,\langle {\bf  x }\rangle) \xi_j(t),
\label{local_langevin_eqs}
\end{equation} 
with the modified deterministic terms given by
\begin{equation} 
\hat a_k({\bf x}^l, \langle {\bf x} \rangle)=\frac{x^l_k(\pi_k({\bf x}^l)-\bar\pi({\bf x}^l)) + \mu (\langle x_k \rangle - x^l_k)}{\bar\pi({\bf x}^l)+\mu}, 
\end{equation} 
where the vector $\langle {\bf x} \rangle=\sum_{l=1}^M {\bf x}^l/M$ with components $\langle x_k \rangle = \sum_{l=1}^M x^l_k/M$ describes the frequencies of the
individual types in the global population and the diffusion terms $\hat c({x}^l,\langle {\bf x} \rangle )$ can be expressed in terms of the modified transition probabilities $\hat T^l_{ik}$ as above.

 Equations (\ref{local_langevin_eqs}) 
 describe the coevolutionary dynamics of the
global population through the coupled evolution of the $\{ {\bf
  x}^1,\dots,{\bf x}^M \}$ local populations. 
In the limit of a large number of local populations ($M \to
\infty$) the distribution of the local populations over the space of
population states (the simplex $S_d$) is described by a density function $\rho({\bf x})$ that is normalized over $S_d$, i.e., $ \int_{S_d}  \rho({\bf x})=1$.
The time evolution of $\rho(\bf x)$ follows a $d-1$ dimensional
advection-diffusion equation -- the Fokker-Planck equation
corresponding to eq. (\ref{local_langevin_eqs}):
\begin{equation}
\dot \rho({\bf x}) =  -\nabla \left\{
 {\bf \hat a}({\bf x}, \langle {\bf x} \rangle) \rho({\bf x}) \right\} + \frac{1}{2}
\nabla^2 \left\{
{\bf \hat b}({\bf x}, \langle {\bf x} \rangle) \rho({\bf x}) \right\},  
\label{adv_diff_eq}
\end{equation}
with the global averages $\langle x_k \rangle = \int_{S_d}  x_k \rho({\bf x})$ coupled back 
in a {\it self-consistent} manner 
into the deterministic terms $ \hat a_k({\bf x}, \langle
      {\bf x} \rangle)$ and the diffusion matrix $ \hat b_{kj}({\bf
        x}, \langle {\bf x} \rangle) = \sum_{i=1}^{d-1} \hat c_{ki}({\bf x}, \langle
     {\bf x} \rangle) \hat c_{ij}({\bf x}, \langle
      {\bf x} \rangle)   $. 
For large local populations ($N \to
\infty $) the diffusion term vanishes as $1/N$.

The above advection-diffusion equation (\ref{adv_diff_eq}) presents an
intuitive picture of the coevolutionary dynamics of the population at a
global scale. We can see that local populations each attempt to follow
the trajectories corresponding to the deterministic replicator
dynamics, while under the influence of two additional opposing
forces: (i) global mixing, which attempts to synchronize local
dynamics and (ii) diffusion resulting from finite population size
effects, which attempts to smear them out over the simplex. The
strength of these forces are tuned by two parameters $\mu$ and $N$,
respectively. 
 
If, further, the effects of synchronization are irrelevant, as for example in the case 
of  populations where selection is externally driven by independent environmental 
fluctuations, we may replace the global population average
 with the time average of any single population. This is the approach we used in our 
study of genetic mixing in bacteria  \cite{SzollosiGEN}. 

During our numerical investigations we found solving the advection-diffusion equation (\ref{adv_diff_eq}) numerically challenging, particularly in the $N\to \infty$ limit. We resorted instead to solving the coupled Langevin equations (\ref{local_langevin_eqs}) for large $M=10^4-10^5$ to simulate the time evolution of $\rho({\bf x})$.
\begin{figure}
\begin{center}
\centerline{\includegraphics[width=.85\textwidth]{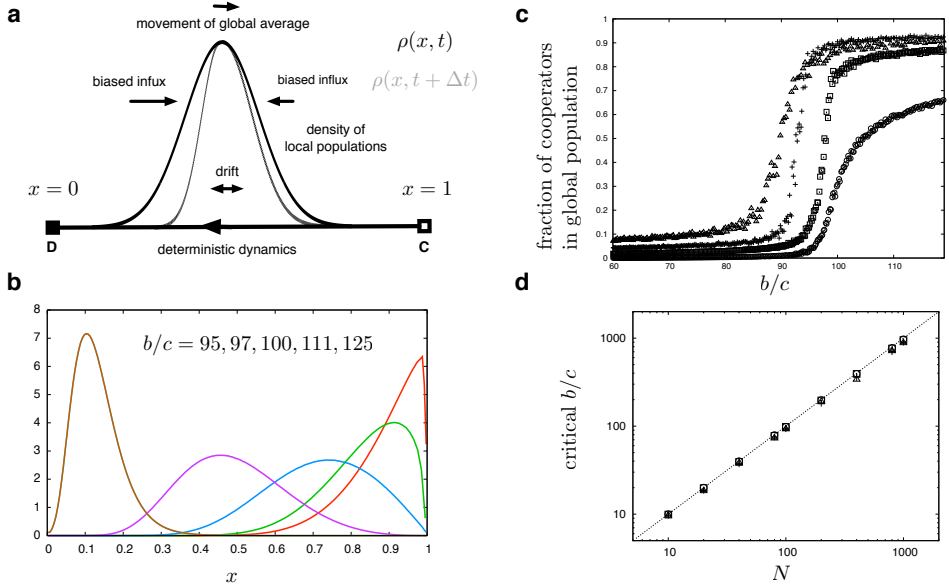}}
\caption{ {\bf a} In an infinitely large  well-mixed population evolutionary dynamics is deterministic and 
leads to the extinction of cooperators as average fitness monotonically declines. The only stable fixed point corresponds to the point where the fraction of cooperators is zero ($x=0$). 
 To understand qualitatively the mechanism favoring cooperation in hierarchically mixed populations  let us  consider some   
density of local populations ($\rho(x,t)$) that is symmetric around its mean at time $t$. Due to
global mixing all local populations are being driven toward the global
average. Due to the influx bias, however, populations with a lower than
average number of cooperators will be driven stronger (faster) than those on
the other side  of the average. Examining the density of local populations at some time $t+\Delta t$, this results in a  net movement of the
global average  toward a  larger fraction of  cooperators. 
This is, of course, opposed  by local reproduction that favors an increase in the
number of  defectors. For the global  average to keep  moving toward a
higher number of cooperators and eventually to keep balance with local
reproduction bias a density of local population with finite width is needed over which the  effect of the influx  bias can 
exert itself.  It is drift  caused by  local  population  size  that
maintains  this  finite width, and  this  is the reason that  the
$b/c$  threshold above which cooperation  dominates
depends on local population size. {\bf b} Stationary density of  local populations $\rho(x)$ for different values of $b/c$ with $N=100$, $\mu=0.1$. {\bf c} Transition toward a global dominance of cooperation for $\mu=10.$ (triangles), $\mu=1$ (crosses), $\mu=0.1$ (squares), $\mu=0.01$ (circles)  with $N=100$. The critical value of $b/c$ depends only weakly on $\mu$ changing by $20 \%$ over four orders of magnitude  {\bf d} Critical values of $b/c$ as a function of $N$ for different values of $\mu$ (notation as before). The dashed line corresponds to $b/c=N$. The critical $b/c$ values were determined by numerically finding the inflection point of the transition curves. $M=10^3$ was used throughout. }
\label{newfig}
\end{center}
\end{figure}

\section{Cooperation in populations with hierarchical levels of mixing}
 The evolution of cooperation is a fundamental problem in biology, as
natural selection under most conditions favors individuals who defect. Despite of this, cooperation is widespread in nature. 
A cooperator is an individual who pays a cost $c$ to provide another individual with some benefit $b$. 
A defector pays no cost and does not distribute any benefits. 
This implies the payoff matrix
\begin{equation}
\begin{pmatrix}
b-c & -c \\
b & 0 \\
\end{pmatrix}, 
\end{equation}
where $b$ is the benefit derived from playing with a cooperator while $c$ is the cost for cooperation.  
From the perspective of evolutionary game theory, which equates payoff with fitness, the apparent dominance of defection is simply the expression of the fact that natural selection \emph{a priori} selects for fitness of individuals and not the fitness of groups.  

 Defection dominates cooperation in any well-mixed
population \cite{HofbauerBOOK}. Population structure induced by spatial structure \cite{NowakNAT92,GTphy} and more general networks of interactions \cite{SantosPRL,OhtsukiNAT06,Csermely}) has, however, been found to facilitate the emergence and maintenance of cooperation. The mechanism responsible, termed spatial, or more generally, network reciprocity\cite{NowakSCI06} depends strongly on the details of local topology. In particular, it seems that lattice like connectivity structures where three-site clique percolation occurs \cite{SzaboClique} and more general interaction graphs where the degree of nodes $k$ does not exceed the ratio of benefit to cost (i.e.\ $k<b/c$) \cite{OhtsukiNAT06} are required for cooperation to be favored. 

Examining the effects of hierarchical mixing on the evolutionary dynamics of cooperation we found that a sharp, but continuous transition leads to the dominance of cooperation as the benefit to cost ratio becomes smaller then the local population size, i.e.\ $b/c < N$. If the benefit to cost ratio is larger then the local population size the global population is dominated by defectors.
The mechanism leading to the dominance of cooperation arises 
due to the competition between local reproduction and global mixing. In local populations with lower average fitness --  larger number of defectors --  the influx of  individuals from the
global scale will be larger  than in local populations with higher average fitness (cf. eq. (6) where the relative strength of the two terms on the left hand side depends on the sum of the average fitness of population $l$ and $\mu$).  The crucial ingredient for cooperation to be successful is population drift 
introduced by  finite  local  population size.  It is  biased influx coupled with  
drift that can result in cooperation  being favored  in  the global  population (Fig \ref{newfig}.). 

   \begin{figure}
\begin{center}
\centerline{\includegraphics[width=.85\textwidth]{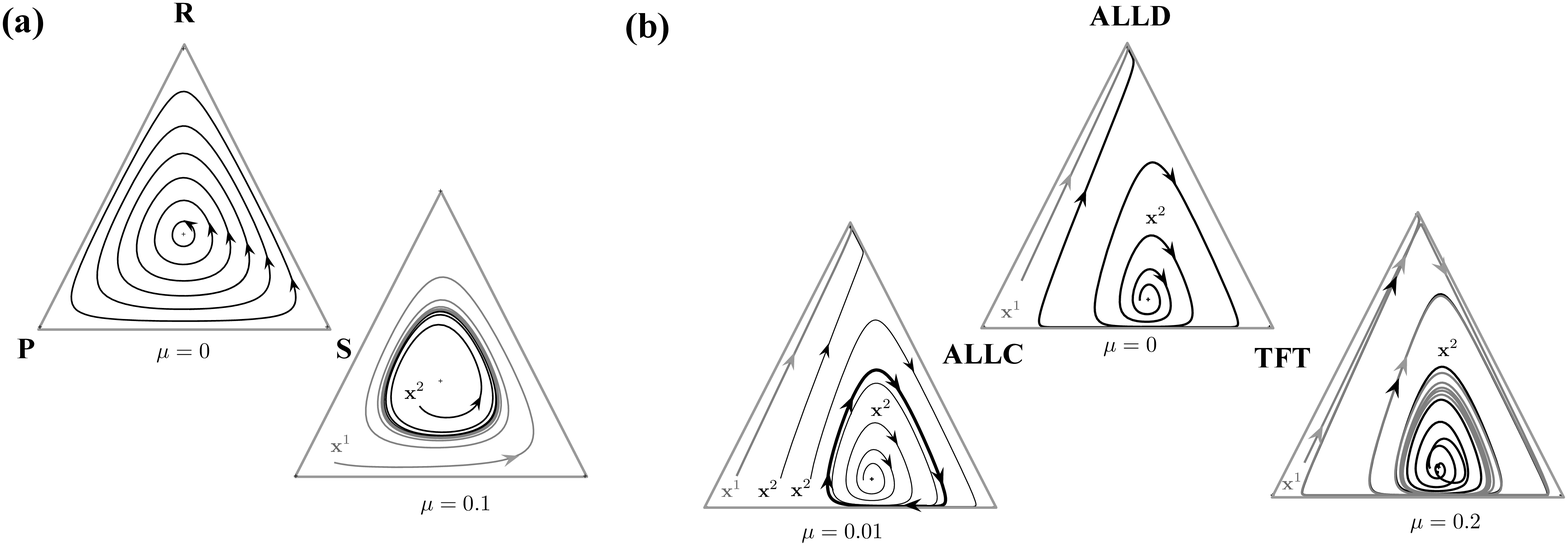}}
\caption{ {\bf a} Deterministic replicator dynamics (the $N \to \infty$
  limit) of the symmetric RPS game consists of neutrally stable orbits
  along which the product of the strategy frequencies $x_{\rm R} x_{\rm P} x_{\rm S}$ is conserved.
 If global mixing is present ($\mu>0$)
  local populations deviate from these neutral orbits toward the global
  average $\langle {\bf x} \rangle$. Considering the simplest system with
  global mixing, that consisting of $M=2$ local populations we see that in the presence of global mixing
  population ${\bf x}^1$ and population ${\bf x}^2$ move toward each
  other, respectively moving closer and further
  from the barycentre of the triangle until they become synchronized and subsequently pursue a common orbit. For deterministic local dynamics ($N\to \infty$) such synchronization invariably occurs for any $M$ if $\mu>0$ and typically converges to the barycentre of the simplex for sufficiently homogeneous initial conditions.   
   {\bf b}
  The deterministic replicator dynamics of the repeated PD game is
  markedly different from that of the RPS game in that the internal fixed point is unstable and in
  the absence of global mixing only ALLD survives. Again turning to
  the simplest scenario with $M=2$ we see that if $\mu=0$ any pair of
 populations ${\bf x}^1$ and ${\bf x}^2$ (gray and black lines) converge to the to the ALLD
corner. As $\mu$ is increased above a critical value a second, stable
configuration emerges: for a large subset of the possible initial
conditions (all, but the left most ${\bf x}^2$) we see that one of the populations (${\bf x}^1$) converges to ALLD
, while the second  (${\bf x}^1$) approaches a limit cycle. If $\mu$ is
increased further, the above configuration ceases to be stable, the
population which initially converges to ALLD (${\bf x}^1$) is subsequently
''pulled out'' by global mixing, following which the two populations
synchronize and are finally absorbed together in ALLD. Simulations,
however, show that synchronization may be avoided for $M>2$ if $\mu$
is not too large.         
}\label{fig2}
\end{center}
\end{figure}

\section{The RPS Game}
To explore the effects of hierarchical mixing in the context of games with three strategies we first turn to the
case of the so called  ''rock-paper-scissors''  (RPS) game.
In the original popular version of the game two players are afforded the chance to
simultaneously display either rock (fist), paper (flat hand) or
scissors (two fingers). If player one displays a flat hand while player two
displays a fist, player one wins as paper wraps rock. Similarly
scissors cut paper, and rocks smashes scissors.  Several examples of
this game have been found in nature  (e.g.\ among lizards  \cite{SinervoNAT}
), but it is bacteria that have
received the most experimental and theoretical attention.

In ecology the often high diversity among microbial organisms in
seemingly uniform environments, referred to as the ''paradox of the
plankton'', has been difficult to understand. Several models based on
spatially explicit game theoretical models have been proposed to
explain this diversity  \cite{CzaranPNAS,LenskiPNASComm,KerrNAT,NowakNATnw}.  These models are all variants of the RPS game
played by colicin producing  bacteria.
Colicins are antibiotics produced by some strains of 
\emph{Echerichia coli}. In experiments (see Fig.1) typically three
strains are used: colicin producing (C), sensitive (S) and resistant
(R). The coevolutionary
dynamics of the three strains can be cast in terms of an RPS game,
C strains kill S strains, but are 
outcompeted, by R strains, because toxin production
involves the suicide of bacteria. The cycle is closed by S strains that
outcompete R strains, because resistance requires mutant versions of
certain membrane protein, which are less efficient than the wild type \cite{KerrNAT}. 
Despite the cyclic dynamics colicin-producing 
strains cannot coexist with
sensitive or resistant strains in a well-mixed culture, yet all three
phenotypes are recovered in natural populations. Local dispersal
(modeled as explicit spatial embedding) has widely been credited with
promoting the maintenance of diversity in this system  \cite{KerrNAT,NowakNATnw,CzaranPNAS,LenskiPNASComm}. 

In its most symmetric form the RPS game is described by
the payoff matrix
\begin{equation}
\begin{pmatrix}
 0 & -\epsilon & \enskip \epsilon \\
 \epsilon & 0 & -\epsilon\\
 -\epsilon & \epsilon & 0 
\end{pmatrix},
\end{equation}
and some $\pi_{\rm base}>\epsilon$.
The dynamics of this game in an infinitely large well mixed population consists of neutral orbits along which 
the product $x_{\rm R} x_{\rm P} x_{\rm S}$ is conserved. For any finite $N$, however, fluctuations
 lead to the inevitable extinction of all but one of the strategies \cite{ReichenbachPRE}.
 Spatial population structure can avert this reduction in diversity \cite{CzaranPNAS,KerrNAT}
 through the emergence of a stable fixed point at the barycentre of the simplex . 
The effect of the gradual randomization of different lattice topologies
 (where a small number of edges are randomly rewired) on the dynamics of the
 game has also been investigated. A Hopf bifurcation leading
 to global oscillations was observed \cite{SzolnokiPRE,SzaboJPA} as the fraction
 of rewired links was increased above some critical value.

\begin{figure}
\begin{center}
\centerline{\includegraphics[width=0.9\textwidth]{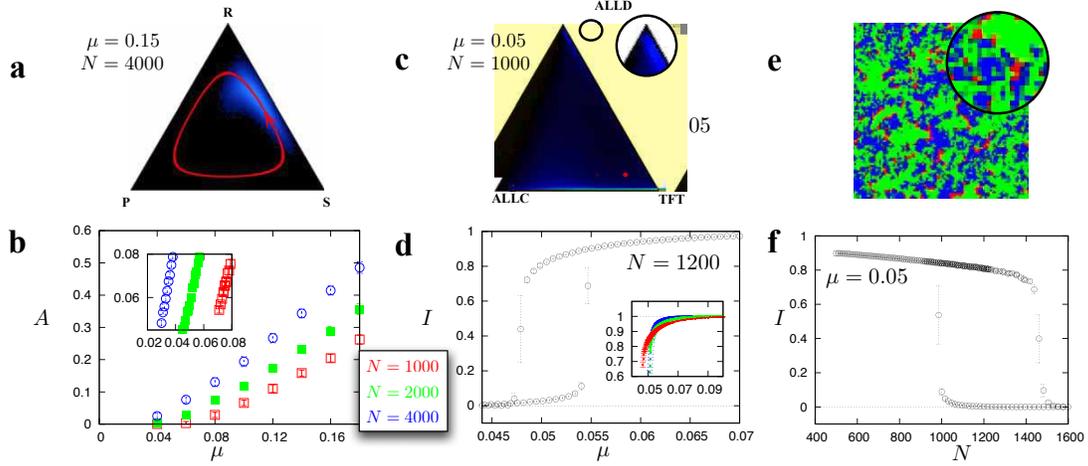}}
\caption{(Color online) {\bf a} In the case of the rock-paper-scissors game a Hopf bifurcation similar 
to that observed for populations evolving on gradually randomized lattices 
 \cite{SzolnokiPRE,SzaboJPA} leads to the emergence of global oscillations 
(the red line indicates the trajectory of $\langle{\bf x }\rangle$) if $\mu$ is larger than a critical value $\mu_{\rm c}(N)$ (see video S1 \cite{EPAPS}). The
 density $\rho({\bf x})$ is indicated with a blue color scale.  {\bf b}  
The ratio $A$ of the area of the global limit cycle and the area of the simplex is plotted as a function of $\mu$ for three different values of $N$. 
 For the repeated prisoner's dilemma game the combination of finite 
local population size and global mixing $\mu>0$ can lead to a stationary 
solution ({\bf c}) qualitatively similar to that observed for explicit spatial embedding {\bf e}. 
This state is characterized by a stable global average (large dot), just as the 
lattice system (data not shown) and sustained local cycles of cooperation, defection
 and reciprocity, also similar to the lattice case where groups of ALLD (red, dark grey) 
individuals are chased by those playing TFT (blue, black), which are gradually outcompeted 
by ALLC (green, light grey). {\bf d}  As $\mu$ is decreased a discontinuous transition can
 be observed to the ALLD phase. The ratio $I$ of populations on the internal cycle is plotted as a function of $\mu$. The inset shows the transition for different values of $N$. {\bf f} The same critical 
line in the $\mu$-$N$ plane can be approached by increasing $N$ with $\mu$ fixed. 
A large hysteresis can be observed as $N$ is decreased below the critical value 
indicating the discontinuous nature of the transition. We numerically simulated  the time evolution of $\rho({\bf x})$ by 
integrating the stochastic differential equation system defined by eq. 
(\ref{local_langevin_eqs}) for large $M$ ($10^4-10^5$) throughout. For the RPS game 
we used  $\pi_{\rm base}=1$ and $\epsilon=0.5$, while in the case of the repeated PD game we followed ref.
 \cite{Imhof}, setting $T=5,R=3,P=1,S=0.1,m=10$ and $c=0.8$. Lattice simulations ({\bf e}) where
 performed on $1000\times1000$ square lattice with an asynchronous local Moran process between
 neighbors and periodic boundary conditions.}
\label{fig3}
\end{center}
\end{figure}

Examining the dynamics of the symmetric RPS game in terms of our hierarchical meanfield
 approximation we observed that an internal fixed point emerged for
 $N\to\infty$ (Fig.3a). 
 More importantly, diversity was also maintained for
 finite local population sizes if global mixing  was present. Simulations of the
 time evolution of $\rho({\bf x})$ also revealed a Hopf bifurcation
 leading to the oscillation of the global average as $\mu$ was increased above a critical
 value $\mu_{\rm c}$ depending on $N$ (Fig.4a). These results show that previous results obtained
 from simulations of populations constrained to different lattice topologies can be
 considered universal in the sense that not only lattices, but any population structure
 that can be approximated by two distinct internally unstructured scales of mixing are sufficient
 for their existence. In the context of the ''paradox of the plankton'' these results imply that
 aside of local dispersal (modeled as explicit spatial embedding) a minimal metapopulation
 structure (with local competition and global migration) can also facilitate the maintenance
 of diversity in cyclic competition systems.
   
\section{The Repeated Prisoner's Dilemma Game}
In the general formulation of the prisoner's dilemma (PD) game, two players have the choice to cooperate or to
defect. Both obtain some payoff $R$ for mutual cooperation and some
lower payoff $P$ for mutual defection. If only one of the players
defects, while the other cooperates, the defector receives the highest
payoff $T$ and the cooperator receives the lowest payoff $S$. That is
$T>R>P>S$ and defection dominates cooperation in any well-mixed
population.
New strategies become possible, however if the game is repeated, and
players are allowed to chose whether to defect or cooperate based on
the previous actions of the opponent. In the following we consider, similar 
to refs.  \cite{NowakNAT04} and  \cite{Imhof} that recently examined the role of finite
 population size and mutation and finite population size, respectively in terms 
of the repeated PD game with three strategies:
 always defect (ALLD), always cooperate (ALLC), and tit-for-tat (TFT).  
TFT cooperates in the first move and then does whatever the opponent did in the previous move.
TFT has been a world champion in the repeated
prisoner's dilemma ever since Axelrod conducted his celebrated computer tournaments
 \cite{AxelrodBOOK}, although it does have weaknesses and may be defeated by
other more complex strategies  \cite{Molander}.

 Previous results indicate that if only the two pure strategies are present 
(players who either always defect or ones who always cooperate) explicit spatial 
embedding  \cite{NowakNAT92} and some sufficiently sparse interaction graphs  \cite{OhtsukiNAT06,Taylor07} 
allow cooperation to survive and the behavior of populations is highly sensitive to the 
underlying topology of the embedding  \cite{SzaboClique}. We have found that introducing global mixing into the PD game with only
the two pure strategies present also allows cooperation to survive. The
mechanism responsible for favoring cooperation in this case, however,
depends on the details of the competition between local reproduction
and global mixing. For more than two strategies these details are much
less relevant and do not qualitatively influence the dynamics. We will,
therefore, consider the delicate issues concerning the PD game with
only the two pure strategies in a separate publication,
and concentrate here on the repeated PD game with three strategies.

\begin{figure}
\begin{center}
\centerline{\includegraphics[width=
.5\textwidth]{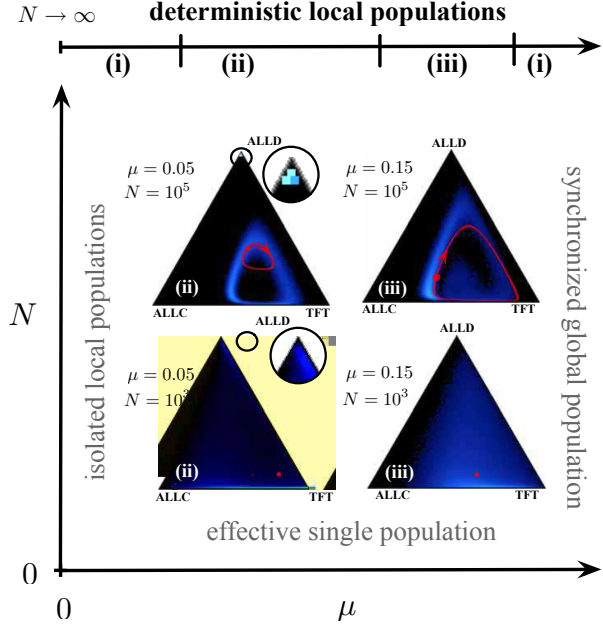}}
\caption{ (Color online) Phase space for the repeated prisoner's dilemma game on a population structure with two distinct scales (see video S2 \cite{EPAPS}).
Three different phases are possible depending on the values of $\mu$ and $N$: (i) only ALLD survives 
(ii) an internal limit cycles is maintained by global mixing due to a large density of 
local populations around the ALLD corner (iii) a globally oscillating
self maintaining limit cycle is formed.
For extreme values of $\mu$ the global dynamics reduces to that of some well-mixed population where only ALLD survives:
As $\mu$ becomes negligible ($\mu \ll \pi_k$ for all $k$) we approach the limit 
of isolated local populations, while for $\mu  \gg \pi_k$ we are left with a single synchronized 
population. Similarly for $N = 2$ -- the smallest system with competition -- the system can
 be described as a single well mixed population for any $\mu$ and ALLD again prevails. In the limit of deterministic local populations ($N\to\infty$) all three phases can be found depending on the value of $\mu$. The
 density $\rho({\bf x})$ is indicated with the color scale.  A figure illustrating the phase space of the repeated prisoner's dilemma game with {\it fitness dependent} global mixing is included in the supplementary material \cite{EPAPS}.}\label{fig4}
\end{center}
\end{figure}

To investigate the effect of global mixing on the repeated PD game with three 
possible strategies: ALLD, ALLC and TFT following 
Imhof {\it et al.\ } \cite{Imhof} we considered the payoff matrix:

\begin{equation}
\begin{pmatrix}
 R m & S m & R m \\
 T m & P m & T + P (m-1)\\
 R m - c & S+P (m-1)-c &  R m -c
\end{pmatrix},
\end{equation}
where the strategies are considered in the order ALLC, ALLD, TFT, 
 $m$ corresponds to the number of 
rounds played and $c$ to the complexity cost associated with conditional strategies (TFT).
The dynamics of this game has a single unstable internal fixed point and the state where each 
member of the population plays ALLD is the only nontrivial stable equilibrium (Fig.3b).      

Introducing global mixing, between local well-mixed populations, however, causes new stationary states to emerge . 
Three phases can be identified: (i)  ALLD wins (ii) large fraction of local populations in
the ALLD corner maintains local cycles of cooperation defection and reciprocity through providing an influx of defectors that prevent TFT players from being outcompeted by ALLC playing individuals (iii) 
a self maintaining internal globally oscillating cycle
emerges. The simplest scenario of two ($M=2$) deterministic ($N\to\infty$) local populations coupled by global mixing ($\mu>0$) already leads to the emergence of phase (ii) as demonstrated in Fig.{\ref{fig2}b} while phase (iii) only emerges for larger $M$. 
For larger $M$ simulations show that in the limit of 
large local populations all global configurations with less than some maximum ratio
of the populations $I$ on the internal cycle are stable in phase (ii). A transition
from phase (ii) to (i) happens as $\mu$ is decreased below a critical
value $\mu^{\rm ii \to i }_{\rm c}$ and $I$ approaches zero as $I
= (1-\mu^{\rm ii \to i}_{\rm c}/\mu)  $ (data not shown). This
can be understood if we considered that near the transition point a
critical proportion $C = \mu (1-I)$ of ALLD individuals needs to
arrive to stabilize local cycles of cooperation defection and
reciprocity. At the critical point $I=0$ and $\mu=\mu^{\rm ii \to i
}_{\rm c}$ which implies $C=\mu^{\rm ii \to i }_{\rm c}$ giving $I
= (1- \mu^{\rm ii \to i }_{\rm c} / \mu)$

Exploring the $N-\mu$ phase space (Fig.5) we see that the
transition from phase (i) to (ii) becomes discontinuous for finite
$N$ (Fig.4d,e). Further, for any given value of $N$ and $\mu$ the global configuration is described by a unique $I$ due to the presence of diffusion. For appropriate values of the parameters the
global average converges to a stationary value in phase (ii)
similarly to case of explicit spatial embedding (Fig.4c). 

For very small ($\mu\ll\pi_k$ for all $k$) and very large ($\mu\ll\pi_k$) values of $\mu$   
 the global dynamics can be reduced
to that of some well-mixed population where only ALLD persists (Fig 5.). For small $N$ we again have an effective well-mixed population -- the only limit were defectors do not dominate is $N\to\infty$. In comparison with
previous results of Imhof {\it et al.} we can see that evolutionary
cycles of cooperation defection and reciprocity can be maintained not
only by mutation, but also by population structures with hierarchical
levels of mixing.

\section{Discussion}

 While it is, of course, clear that the reduction of any realistic 
population structure to a manageable construction is always an
approximation, it has not been clearly established what the relevant
degrees of freedom are in terms of evolutionary dynamics.
Meanfield approximations are a classic method of statistical and
condensed matter physics and are routinely used to circumvent intractable
combinatorial problems which arise in many-body systems. 
Cluster-meanfield approximations of
sufficient precision \cite{GTphy,Hui07} have been developed that adequately
describe the evolutionary dynamics of explicitly structured populations 
through systematically approximating the
combinatorial complexity of the entire topology with that of small
motif of appropriate symmetry. The effects of more minimal effective
topologies have, however, not been investigated previously.
In the above we have shown that straightforward 
hierarchical application of the meanfield approximation (the
assumption of a well-mixed system) surprisingly unveils a new level of
complexity.

In the broader context of ecological and population genetics research on structured populations our model can be described as a metapopulation model. The term 'metapopulation' is, however, often used for any spatially structured population \cite{Hanski}, and models thereof. More restrictive definitions of the term are often implied in the context of ecology and population genetics literature.  

The foundations of the classic metapopulation concept where laid down by Levin's vision of a ''metapopulation" as a population of ephemeral local populations prone to extinction. A classic metapopulation persists, like an ordinary population of mortal individuals, in a balance between 'deaths'  (local extinctions) and 'births' (establishment of new populations at unoccupied sites)  \cite{Hanski}. This classic framework is most wide spread in the ecology literature, a less often employed extension is the concept of a structured metapopulation where the state of the individual populations is considered in more detail, this is more similar to our concept of hierarchical mixing, but differs in considering the possibility of local extinctions.      

The effects of finite population size and migration, which our model considers, has been of more central concern in the population genetics literature. The analog of Levin's classic metapopulation concept is often referred to as the 'finite-island' model \cite{Pannell} the effective population genetic parameters describing which, have been explored in detail\cite{whitlock}. The study of the population genetics of spatially subdivided populations in fact predates Levin, Wright having emphasised the capacity of drift in small populations to bring about genetic differentiation in the face of selection and/or migration several decades prior\cite{Pannell}.      

Our hierarchical mixing model treats the coevolutionary dynamics of evolutionary games on structured populations in a manner similar to the most simple population genetic models of spatially subdivided populations, focusing on the parallel effects of selection, drift and migration. It goes beyond these models both in considering the effects of frequency dependent selection (and the strategic aspects of the evolutionary dynamics this implies) and in using a self-consistent approach to describe the global state of the subdivided population. Also, in order to maintain a connection with previous work on the effects of spatial structure on evolutionary games, which rely on Nowak's concept of spatial games \cite{NowakNAT92}, with individuals restricted to interact, and hence compete, only with neighbours as defined by some topology of interaction, we develop our model from the level of the individual by introducing a modified version of the Moran process -- and not by extending  the Wright-Fisher process (which considers discrete generations and binomial sampling to account for finite population size). The effective population structure described by our hierarchical mixing model can be thought of as a population of individuals, interactions among which are specified by the edges of a hierarchically organized random graph. The fundamental difference in our picture is that the edges of this graph of interactions are not considered to be fixed, but are instead in a constant state of change, being present with a different probability between pairs of individuals who share the same local population and between pairs of individuals who do not (Fig.1.). We consider annealed randomness, which in contrast to the usual quenched picture of fixed edges is insensitive to the details of topology. Our approach we believe best facilitates the exploration of the effects of changing the relative strengths of drift and migration in the context of evolutionary games on structured populations.

Examining the effects of hierarchical mixing in the context of the evolution of robustness we demonstrated that biased influx coupled with drift can result in cooperation being favored, provided the ratio of benefit to cost exceeds the local population size. This result bears striking resemblance to that of Ohtsuki {\it et al.} \cite{OhtsukiNAT06}, who were able to calculate the fixation probability of a randomly placed  mutant for any two-person, two-strategy game on a regular graph and found that cooperation is favored provided the ratio of benefit to cost exceeds the degree of the graph. Our results demonstrate that this rule extends to the minimal spatial structure induced by hierarchical levels of mixing.

Applying our model of spatial structure to the repeated prisoners dilemma revealed that a constant influx of defectors can help to stabilize cycles of cooperation, defection, and reciprocity through preventing the emergence of an intermittent period of ALLC domination in the population, which would present a situation that ''leaves the door wide open'' to domination by defectors. While previous work has been done on the effects of ''forcing'' cooperation  \cite{SzaboForcing} the idea that an influx of defectors can in fact stabilize the role of reciprocity in promoting cooperation has not been proposed previously. 
 It seems highly unlikely that this mechanism can be explained in terms of kin or multilevel (group) selection, the similarities between which in structured populations have recently been the subject of intensive debate (see e.g.\ \cite{Killingback} and \cite{Grafen}  or \cite{TraulsenML} and \cite{Lehmann}). Kin selection can operate whenever interactions occurring among individuals who share a more recent common ancestor than individuals sampled randomly from the whole population \cite{Lehmann} are relevant. In our case it is the interaction between defectors, arriving from the global scale, and TFT players present at the local scale that is important, and not the interaction between individuals in the local population, who may be thought of as sharing a recent common ancestor due to local dispersal. Also, while the concept of multilevel selection presents a promising framework for the study of evolution of cooperation, it must nonetheless be possible to derive it from ''first principles'' -- just as kin selection can be cast as an emergent effect of local dispersal.

While there has been considerable work on studying the evolutionary games on graphs and highly symmetric spatial structures very little attention has been paid to the effects of more minimal effective population structures, despite their widespread application in ecology and population genetics, fields from which evolutionary game theory was born and must ultimately reconnect with.       
We believe that the minimal 
population structure that such a hierarchical meanfield theory describes is potentially more relevant in a wide range of
natural systems, than more subtle setups with a delicate
dependence on the details and symmetries of the topology. We showed through two examples
that such structure is sufficient for the emergence
of some phenomena previously only observed for explicit spatial
embedding, demonstrating the potential of our model to identify robust  effects of population
structure on the dynamics of evolutionary games that do not depend on
the details of the underlying topology.
The practical advantage of our approach, lies in its ability to readily determine whether or not some feature of a structured population
 depends on the topological details of local interactions.

Recent simulation result concerning the dynamics of public goods games on
different population structures  \cite{HauertSCI02,HauertPRL02} and
experiments where global mixing in an RPS like bacteria-phage system lead to
the emergence of a ''Tragedy of the commons'' scenario  \cite{KerrTCNAT} should all be
 amicable to analysis in terms of our method.

\section{Acknowledgments}

 This work was partially supported by the Hungarian Scientific Research 
Fund under grant No: OTKA 60665.

\section{Appendix}

Our approach readily generalizes for an arbitrary number of hierarchical mixing levels. For three levels of mixing we may consider the global population to be comprised of $\mathcal{M}$ subpopulations each of which is in turn subdivided into $M$ local populations. With $m\in\{1,\cdots,\mathcal{M}\}$ running over subpopulations and $l\in\{1,\cdots,M\}$ over local populations the transition probabilities can be written as:  
\begin{equation}
\hat T^{ml}_{ik} = \frac{n^{ml}_i}{N} \left( \frac{\pi^{ml}_k n_k^{ml} + \mu^{(1)} \langle ^{(1)}\sigma_k^{ml'}\rangle_{l'} + \mu^{(2)} \langle\langle ^{(2)}\sigma_k^{m'l'}\rangle_{l'}\rangle_{m'} }{\sum_{k=1}^{d}(\pi^{ml}_k n_k^{ml} + \mu^{(1)} \langle ^{(1)}\sigma_k^{ml'}\rangle_{l'} + \mu^{(2)} \langle\langle ^{(2)}\sigma_k^{m'l'}\rangle_{l'}\rangle_{m'})}\right),
\end{equation}
where primed indices indicate the scale of mixing over which the average is taken, $\mu^{(1)}$ describes the strength of mixing, and the $^{(1)}\sigma_k^{ml}$ the tendencies of mixing among local populations within a subpopulation, while  $\mu^{(2)}$ describes the strength of mixing, and the $^{(2)}\sigma_k^{ml}$ the tendencies of mixing among subpopulations in the global population.

\end{document}